\documentclass[twocolumn,english,prl]{revtex4}
\usepackage[T1]{fontenc}
\usepackage[latin9]{inputenc}
\usepackage{float}
\usepackage{amsmath}
\usepackage{amssymb}
\usepackage{graphicx}
\usepackage{esint}
\usepackage{color}

\makeatletter
\@ifundefined{textcolor}{}
{%
 \definecolor{BLACK}{gray}{0}
 \definecolor{WHITE}{gray}{1}
 \definecolor{RED}{rgb}{1,0,0}
 \definecolor{GREEN}{rgb}{0,1,0}
 \definecolor{BLUE}{rgb}{0,0,1}
 \definecolor{CYAN}{cmyk}{1,0,0,0}
 \definecolor{MAGENTA}{cmyk}{0,1,0,0}
 \definecolor{YELLOW}{cmyk}{0,0,1,0}
}

\usepackage{babel}

\usepackage{babel}

\makeatother

\usepackage{babel}
\begin{document}

\title{Engineering Quantum Anomalous Hall Plateaus\\ and Anti-Chiral States with AC Fields}

\author{\'{A}lvaro G\'{o}mez-Le\'{o}n$^{1}$, Pierre Delplace$^{2}$, and Gloria Platero$^{1}$}

\affiliation{$^{1}$Instituto de Ciencia de Materiales, CSIC, Cantoblanco, Madrid
E-28049, Spain.\\
 $^{2}$D\'{e}partement de Physique Th\'eorique, Universit\'e de Gen\`{e}ve, CH-1211
Gen\`{e}ve, Switzerland.}

\date{\today}
\begin{abstract}
We investigate the AC electric field induced quantum anomalous Hall effect in honeycomb lattices
and derive the full phase diagram for arbitrary field amplitude and phase polarization. 
We show how to induce anti-chiral edge modes as well as topological phases characterized by a Chern number larger than $1$ by means of suitable drivings.
In particular, we find that the Chern number develops plateaus as a function of the frequency, 
providing an time-dependent analogue to the ones in the quantum Hall effect. 
\end{abstract}
\maketitle
\textbf{\textit{Introduction:}} The realization
of different topological states of matter is one of the major challenges
for both fundamental reasons and technological perspectives. Several
of these states have been originally predicted in the honeycomb lattice, whose Dirac-like
band structure brings the system at a critical point of a topological
phase transition. There, the development of gaps by different mechanisms results in a variety of 
topological phases \cite{HaldaneModel,Kane-Mele}. In this line, the quantum spin Hall phase
was first obtained in HgTe quantum wells \cite{DiscoveryTI1,DiscoveryTI},
where the topological phase transition was controlled by the thickness of the quantum well.
Likewise, the quantum anomalous Hall effect (QAHE) \textendash{} for which
time-reversal symmetry (TRS) is broken in the absence of a magnetic
field \textendash{} was originally proposed
by Haldane in the honeycomb lattice \cite{HaldaneModel}. However, 
despite the success of the theoretical model, it has been very difficult to achieve experimentally,
until very recently in doped topological insulators \cite{ExperimentAnomalous}.

Simultaneously, different techniques have been proposed to externally control
the topological properties of a system. One of the most promising
consists in periodically drive the system in order to achieve a topological phase transition\cite{Tanaka,Galiski,AGL,RUdnerPRX,DEmlerPRB,Cirac-Driven-Majoranas}.
Proposals with a time periodic driving
in semiconductors \cite{Dora,Galiski}, optical lattices\cite{HAuke} or
graphene \cite{Tanaka,FertigPAper,DemlerPRBGRaphene,Fito,KAtan-Graphene,TorresPRB} have been suggested
to achieve various dynamical generalizations of static topological
phases, called \textit{Floquet} topological phases \cite{Dora}, and have been recently observed in photonic crystals \cite{rechtsman13}. 
Importantly for our purposes, the previous studies on the honeycomb lattice were restricted either to full numerical calculations, very high
frequencies and/or to a low energy approximation, thus constraining
the value of the Chern number to $0,\pm1$ and leaving unknown the main part of
the phase diagram.

In the present work, we derive the full phase diagram of periodically-driven honeycomb lattices in the QAHE regime,
by explicitly calculating the Chern number of the Floquet
bands. Our model is valid for arbitrary field frequency, amplitude and polarization and therefore goes beyond the single Dirac cone description
and the rotating wave approximation.

Surprisingly, we find that a clockwise driving may also lead to counterclockwise
(or anti-chiral) edge states. Moreover, we show how to induce topological phases with Chern numbers larger than $1$,
and we discuss the emergence of new chiral and anti-chiral edge states
for different boundaries. Finally, we find that the
Chern number, as a function of the frequency, develops a plateau structure
which provides a Floquet analogue of the quantum Hall effect plateaus.

\textbf{\textit{Model:}} We consider non-interacting spinless particles in a honeycomb lattice, coupled to an in-plane, time-dependent, 
spatially homogeneous vector potential $\mathbf{A}\left(\tau\right)$ of period $T=\frac{2\pi}{\omega}$ 
and formulate the problem within the Floquet formalism.
In general, the hopping parameters to the $j$ first neighbors in the presence of an AC field read $t_{j}\left(\tau\right)=te^{i\mathbf{A}\left(\tau\right)\cdot\mathbf{d}_{j}}$.
Due to the time and spatial periodicities, the system
is described by Floquet-Bloch states which fulfill the Floquet eigenvalue
equation: $\mathcal{H}\left(\mathbf{k},\tau\right)|u_{\alpha,\mathbf{k}}\left(\tau\right)\rangle=\epsilon_{\alpha,\mathbf{k}}|u_{\alpha,\mathbf{k}}\left(\tau\right)\rangle$,
where $\mathcal{H}\left(\mathbf{k},\tau\right)\equiv H\left(\mathbf{k},\tau\right)-i\partial_{\tau}$
is the Floquet operator, $H\left(\mathbf{k},\tau\right)$ the time
dependent Bloch Hamiltonian, $\alpha$ the Floquet band index, $|u_{\alpha,\mathbf{k}}\left(\tau\right)\rangle$
the $T-$periodic Floquet-Bloch states and $\epsilon_{\alpha,\mathbf{k}}$
the quasi-energy which is defined modulo $\omega$. It is useful to
decompose $\mathcal{H}\left(\mathbf{k},\tau\right)$ in Fourier components
in time to calculate the quasi-energies. They read (see supplementary material): 
\begin{eqnarray}
\langle\langle u_{\alpha,\mathbf{k},p^{\prime}}|\mathcal{H}\left(\mathbf{k},\tau\right)|u_{\beta,\mathbf{k},p}\rangle\rangle & = & \tilde{t}_{p^{\prime},p}^{\alpha,\beta}-p\omega\delta_{p^{\prime},p}\delta_{\alpha,\beta},\label{eq:FH-FourierSapce}
\end{eqnarray}
where the hoppings between Fourier components $\tilde{t}_{p^{\prime},p}^{\alpha,\beta}$
are given in terms of the undriven hoppings: 
\begin{equation}
\tilde{t}_{p^{\prime},p}^{\alpha,\beta}\equiv\frac{1}{T}\int_{0}^{T}\sum_{j}e^{i\omega\tau\left(p^{\prime}-p\right)}e^{i\mathbf{k}\cdot\mathbf{d}_{j}}t_{j}^{\alpha,\beta}\left(\tau\right)d\tau\ .\label{eq:Hoppings-FourierSpace}
\end{equation}
For a honeycomb lattice $\mathbf{d}_{1}=a\left(1,0\right)$,
$\mathbf{d}_{2}=a\left(-\frac{1}{2},\frac{\sqrt{3}}{2}\right)$, $\mathbf{d}_{3}=a\left(-\frac{1}{2},-\frac{\sqrt{3}}{2}\right)$
with $a$ the inter-site spacing. Then the Fourier components of the
time dependent hoppings are ($q=p^{\prime}-p$): 
\begin{equation}
\tilde{t}_{p^{\prime},p}=H_{q}\left(\mathbf{k}\right)=\left(\begin{array}{cc}
0 & \rho_{q}\left(\mathbf{k}\right)\\
\rho_{-q}^{*}\left(\mathbf{k}\right) & 0
\end{array}\right),\label{eq:FourierHam}
\end{equation}
with $\rho_{q}(\mathbf{k})=\sum_{j}t_{j,q}^{F}e^{i\mathbf{k}\cdot\mathbf{a}_{j}}$,
$\mathbf{a}_{1}=0$, $\mathbf{a}_{2}=a\left(\frac{3}{2},\frac{\sqrt{3}}{2}\right)$,
$\mathbf{a}_{3}=a\left(\frac{3}{2},-\frac{\sqrt{3}}{2}\right)$. Finally, for a vector potential of the form
$\mathbf{A}\left(\tau\right)=\left(A_{x}\sin\left(\omega\tau\right),A_{y}\sin\left(\omega\tau+\phi\right),0\right)$,
with $\phi$ the phase difference which takes into account the in-plane rotation of the field, 
the renormalized \textit{dressed} hoppings read $t_{j,q}^{F}=te^{iq\Psi_{j}}J_{q}\left(\mathcal{A}_{j}\right)$
\cite{OURPAPER}, where the functions $\Psi_{j}$ and $\mathcal{A}_{j}$
encode all the information of the AC field configuration, with $\Psi_{1}=0$
and: 
\begin{eqnarray}
\tan \Psi_{2,3} & = & \mp\frac{\sqrt{3}A_{y}\sin\left(\phi\right)}{A_{x}\mp\sqrt{3}A_{y}\cos\left(\phi\right)}\label{eq:phase}\\
\mathcal{A}_{1}=A_{x}a,\ \mathcal{A}_{2,3} & = & \frac{a}{2}\sqrt{A_{x}^{2}+3A_{y}^{2}\mp2\sqrt{3}A_{x}A_{y}\cos\left(\phi\right)}\ .\nonumber 
\end{eqnarray}
We stress that the previous derivation is  not restricted to a single Dirac cone and is valid for any field polarization, frequency and intensity.

 In physical realizations of honeycomb systems, additional Bloch bands are present as for instance
the sigma bands in graphene, the higher energy modes in the wave guides of photonic crystals or in the harmonic confinement potential of the trapped cold atoms.
These additional bands are specific to each realization of honeycomb lattices and their treatment is out of the scope of the present work.


\textbf{\textit{Phase diagram of the lattice model:}} In the high
frequency regime ($\omega\gg t$) the Floquet operator (Eq.\ref{eq:FH-FourierSapce})
is approximately block diagonal, meaning that the absorption/emission of photons with energy $\omega$
is very unlikely. Then, one can choose a single block as the effective
Floquet operator (note that all blocks are equivalent, and related
by a shift $p\omega$, with $p$ an integer). In this limit, the system fulfills both time-reversal
 and particle-hole symmetries, and its Hamiltonian is analogous
to the one of an undriven honeycomb lattice with renormalized anisotropic hoppings
$t_{j,0}^{F}=tJ_{0}\left(\mathcal{A}_{j}\right)$. The hopping anisotropy
breaks $C_3$ symmetry and changes the position of the
Dirac points. When the renormalized hoppings fulfill $\pm t_{i,0}^{F}\pm t_{j,0}^{F} \pm t_{k,0}^{F}=0$
($i\neq j\neq k$), the Dirac points merge, leading to an insulating
state\cite{OURPAPER}. Fig.\ref{fig:Phase-diagram} (Top) shows, for $A_{x}=A_{y}$, the phase diagram of several Dirac semi-metalic (SM) and Zak insulating (ZI) 
phases which exhibit non dispersive zero energy edge modes (see the supplementary material).
\begin{figure}
\includegraphics[scale=0.35]{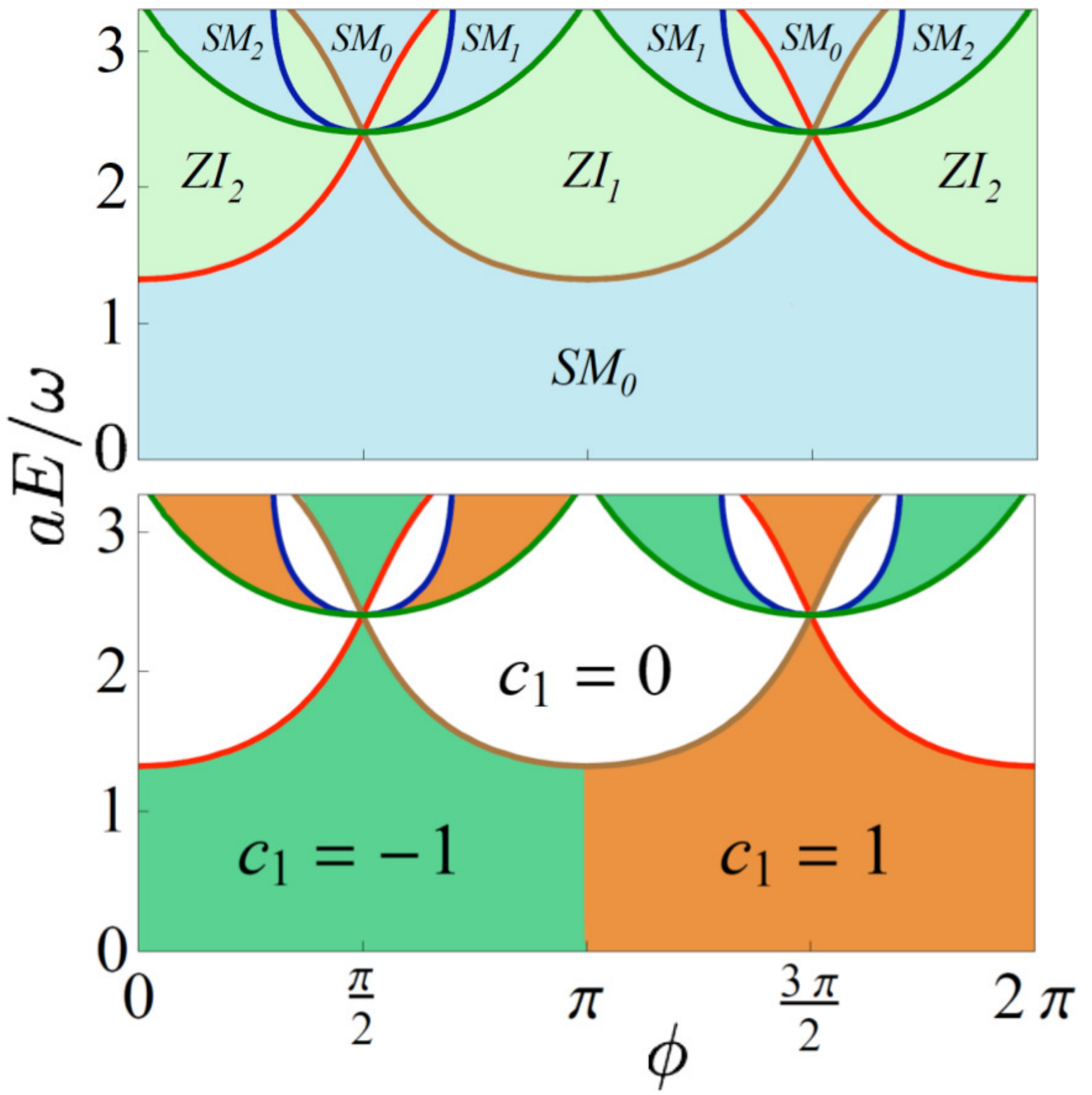} \caption{\label{fig:Phase-diagram} (Top) Phase diagram obtained for the unperturbed Hamiltonian $H_0$ showing semi-metallic $SM_{i}$
(light blue) and insulating phases $ZI_{i}$ (light green). They are obtained
by the merging of the Dirac points at the different time reversal symmetric
points of the first Brillouin zone. (Bottom) First
Chern number values at the lower Floquet band for $\omega=10t$, when first
order coupling among the side-bands is considered (green: $c_{1}$=-1, white:  $c_{1}$=0; orange:  $c_{1}$=1). 
}
\end{figure}
However, even at high frequency, a small coupling between the
Floquet side-bands exists. If this coupling breaks TRS \textendash{}
as for the case of non-linearly polarized fields \textendash{} an
\textit{inverted} gap opens\cite{DemlerPRBGRaphene}. This can be understood
in terms of the complex phases $\Psi_{i}$ attached to the renormalized hoppings
for non-linearly polarized fields. In order to take into
account the effect of this small coupling, we now use first order
perturbation theory in $t/\omega$ and derive an effective time independent
Bloch Hamiltonian for the steady state (details in the supplementary material): 
\begin{equation}
H_{\text{eff}}(\mathbf{k})=H_{0}-\frac{1}{\omega}\left(\left[H_{0},H_{-1}\right]-\left[H_{0},H_{1}\right]+\left[H_{-1},H_{1}\right]\right)\ .\label{eq:Effective1}
\end{equation}
The correction to $H_{0}(\mathbf{k})$ is proportional to $t^2\sigma_{z}/\omega$, where $\sigma_z$ is the Pauli matrix,
and opens a quasi-energy \textit{dynamical} gap at the Dirac points
for $\phi\neq0,\pi$.
The k-dependence of this correction yields a valley dependent sign of the driving-induced \textit{mass term}. 
A low $k$ expansion of this expression gives the results found in \cite{DemlerPRBGRaphene,CaysolReview,Torres,KAtan-Graphene}.

We now characterize the topology of the AC driven
lattice by explicitly calculating the Chern number from $H_{\text{eff}}(\mathbf{k})=\mathbf{h}(\mathbf{k})\cdot \vec{\sigma}$ 
where $\vec{\sigma}$ denotes the vector of Pauli matrices.
Following the elegant method of the Brouwer degree developed in Ref.\cite{Breuwerdegree}, the Chern number
(of the lower quasi-energy band) reads:
\begin{equation}
c_1=\frac{1}{2} \sum_{\mathbf{D}_i} \text{sgn}\left[\partial_{k_x}h_x\partial_{k_y}h_y - \partial_{k_x}h_y\partial_{k_y}h_x \right]_i \text{sgn}\left[h_z\right]_i
\label{chern}
\end{equation}
This analytical approach requires to know the position of the Dirac points $\mathbf{D}_i$ for any value of
the field, which is obtained from the expression of the dressed
hoppings $t_{j,0}^{F}(A_{x},A_{y},\phi)$ (Eq.\ref{eq:phase}).

We show in Fig.\ref{fig:Phase-diagram} (bottom) the values of the Chern number for $A_{x}=A_{y}$ in the high frequency regime considering the effective Hamiltonian described above where the coupling between the side bands up to first order in perturbation theory has been included.
This phase diagram is the first main result of the present work. One can realize that while the Chern number remains zero for the insulating phases, 
it becomes $\pm1$ in the perturbed SM phases. 
Importantly, our results also show that a change of the Chern number can originate from two different mechanisms:
The first mechanism corresponds to the usual Haldane-like gap opening by TRS breaking \cite{HaldaneModel} encoded into a mass term
of opposite sign for the two Dirac cones. This is consistent with previous studies working within the single Dirac cone approximation and restricted to weak amplitudes of the driving \cite{DemlerPRBGRaphene,Dora}. 
It follows from this mechanism that a change of the chirality of the field changes the sign of the mass term $h_z$ in Eq.\ref{chern}, thus reversing the sign of the Chern number.
In contrast, the second mechanism corresponds to a change of chirality for each Dirac cone (first factor in Eq.\ref{chern}). 
This happens when either the intensity or the polarization of the field is continuously varied so that a pair of Dirac cones is successively annihilated 
and created at two different points of the Brillouin zone, thus reversing their vorticity \cite{OURPAPER}. 
Therefore, the field polarization does not fix by itself the value of the Chern number.
This leads to the counterintuitive result that a clockwise driving field can give rise to a counter clockwise edge mode, or \textit{anti-chiral} edge state.
To illustrate this mechanism, we show in the supplementary material the quasi-energy spectra for ribbons with different boundaries.

\textbf{\textit{Multi-photon resonances and emergence of plateaus:}}
We now ask whether the driving can induce gapped phases with
$|c_1|>1$. One way to address this question is 
to notice that in order to obtain larger Chern numbers, it is necessary to
have more than one pair of Dirac points \cite{Breuwerdegree}. Previous studies have shown
that multi-photon resonances, occurring when the Floquet bands overlap,
can induce additional pairs of Dirac points, provided that TRS holds
\cite{OURPAPER,Galitski-cones}. Such an overlap can be achieved by
decreasing the frequency, which brings the Floquet side-bands closer to each other.
It is therefore natural to investigate the effect of the frequency
decrease on the topology of the system for drivings that break TRS.
However, a decrease in frequency increases the number of Floquet bands involved
in the description of the system, and as one approaches $t\sim\omega$,
the validity of $H_{\text{eff}}$ breaks down. In this regime the dynamics becomes highly non-linear, and a numerical treatment
is required. 

Let us first describe how the system is affected when
the side-bands overlap. In periodically driven systems, we must distinguish
two inequivalent gaps: the one between the conduction and the valence
bands \textit{within} the same Floquet side-band $\Delta_{0}$, and
the one separating two \textit{different} side-bands $\Delta_{\pi}$,
as depicted in Fig.\ref{fig:Schematic}.
Note that in the high frequency limit, the $\Delta_{\pi}$ gap was assumed to be infinite: 
this corresponds to the "atomic" limit between side-bands.
\begin{figure}
\includegraphics[scale=0.55]{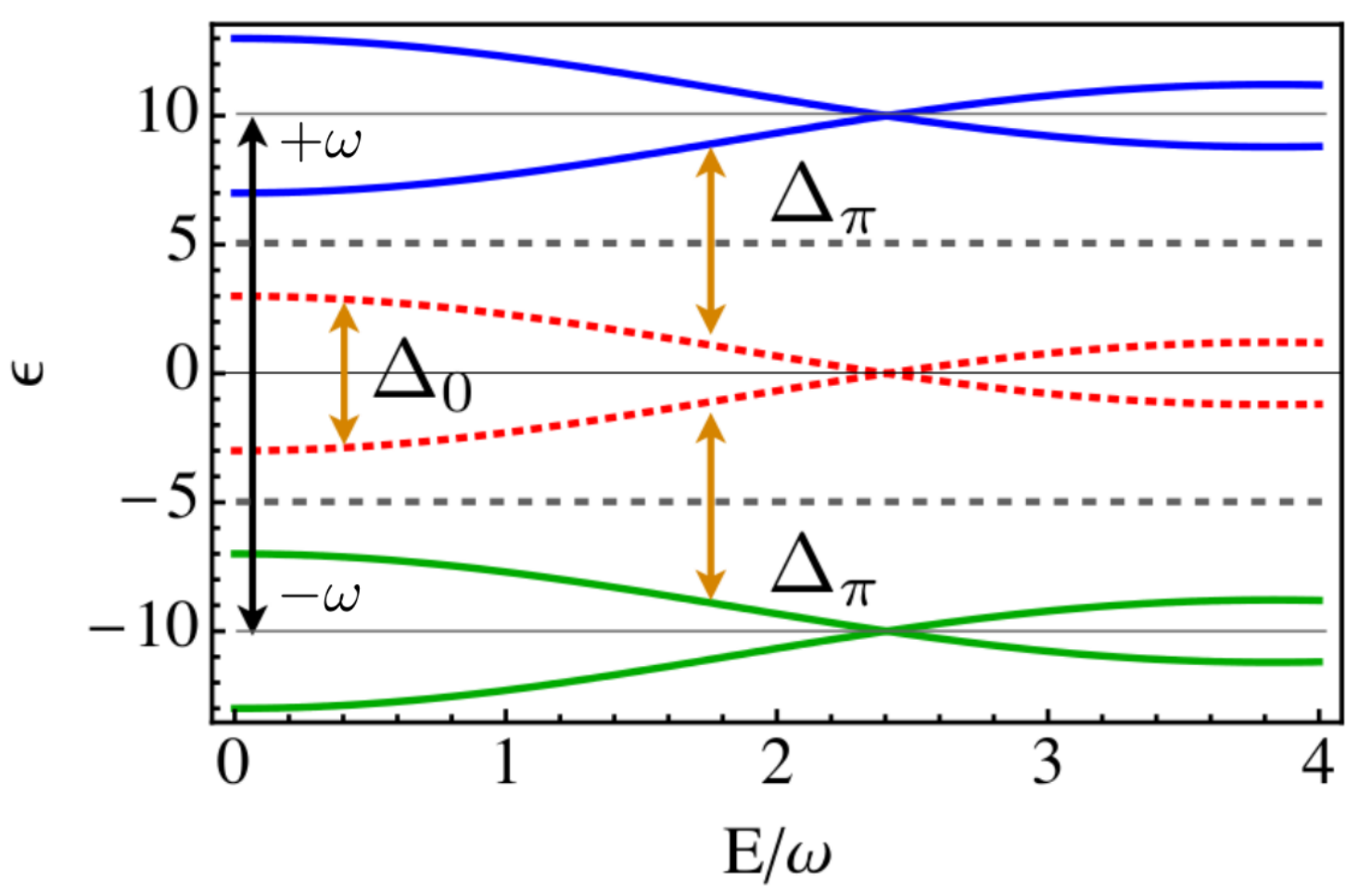}
\caption{\label{fig:Schematic}Schematic representation of three equivalent
side-bands separated by a $\omega$ shift, and the two inequivalent gaps $\Delta_{0,\pi}$.}
\end{figure}
When neighboring side-bands touch due to a frequency decrease, the
gap $\Delta_{\pi}$ closes and a topological phase transition can
happen. 

Fig.\ref{fig:edgestates} (a-b) shows the quasi-energy spectra
for zigzag and armchair ribbons after the gap $\Delta_{\pi}$ has
closed and re-opened by decreasing the frequency up to $\omega=2.5t$.
There, two chiral edge states with opposite chirality to the pre-existing
one in the gap $\Delta_{0}$ have emerged along the same boundary. 
Thus, the total chirality has changed sign whereas the field polarization has not. 
The system can therefore support chiral and anti-chiral edge states simultaneously.
Importantly, the quasi-energy shift of $\pm\omega/2$, makes the
states in the gap $\Delta_{\pi}$ different from the ones in the gap
$\Delta_{0}$ since they contain a time dependent correlation among the spinor components \cite{Cirac-Driven-Majoranas,Transport-Signatures,Torres}.
It is therefore important to know whether two edge states with opposite
chirality lie in the same gap or not.
The Chern number of this phase can be deduced from the relation $c_{1}=W_{0}-W_{\pi}=3$
where $W_{0/\pi}$ is the total chirality of the edge states lying
in the gap $\Delta_{0/\pi}$\cite{RUdnerPRX}.
Fig.\ref{fig:edgestates} (c-d) shows the previous ribbons spectra for $\omega=1.7t$, after both $\Delta_{0,\pi}$
 have closed and re-opened.
 Note that the number of edge states crossing each gap depends on the geometry of the edge,
 although the total chirality in each gap ($W_{0,\pi}=0,4$ respectively) does not. 
This illustrates that $W_{0,\pi}$ is another topological invariant of the system.
It can actually be expressed as a 3D winding number in the extended $(\mathbf{k},t)$ space \cite{RUdnerPRX}.

\begin{figure}[h!]
\includegraphics[scale=0.32]{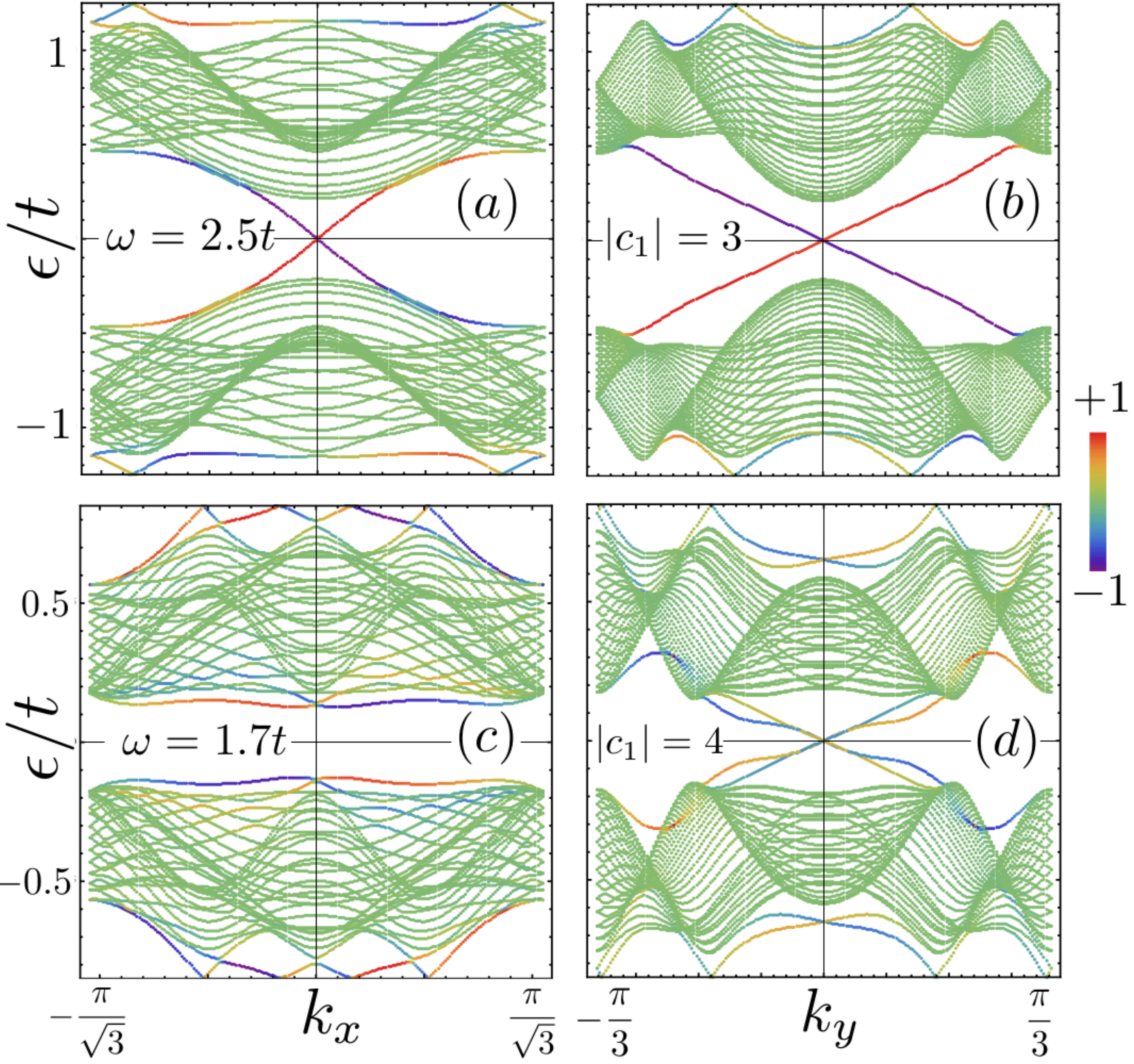} 
\caption{\label{fig:edgestates}Quasi-energy spectra for
zigzag (b,d), and armchair (a,c) ribbons at $\omega=2.5t$
(top) and $\omega=1.7t$ (bottom) respectively. The field amplitude
and polarization are $A_{x}a=A_{y}a=1$, and $\phi=\pi/2$. Chiral edge
states connect the valence and the conduction band. They are localized 
at the edges according to the time average density operator
 $\bar{\rho}(\mathbf{r}_0)-\bar{\rho}(\mathbf{r}_N)$ plotted in color code.} 
\end{figure}
 Next, we use a numerical method\cite{Hatsugai-NumericalChern}
to explicitly compute $c_1$ for arbitrary $\omega$ and plot its (absolute) value in Fig.\ref{fig:Plateaus}. 
This is the second main result of this work. It shows a plateau structure with oscillations whose period decreases with $\omega$. 
The oscillations are due to the coexistence of chiral and anti-chiral states, and reflect the dynamical topological phase transitions induced by the alternating closures and re-openings of the gaps $\Delta_{0,\pi}$ 
as the frequency is tuned. Importantly, for a
given plateau, it is still possible to change $c_1$ by
tuning the amplitude or the polarization of the driving, as in Fig.\ref{fig:Phase-diagram}.
\begin{figure}[h!]
\includegraphics[scale=0.26]{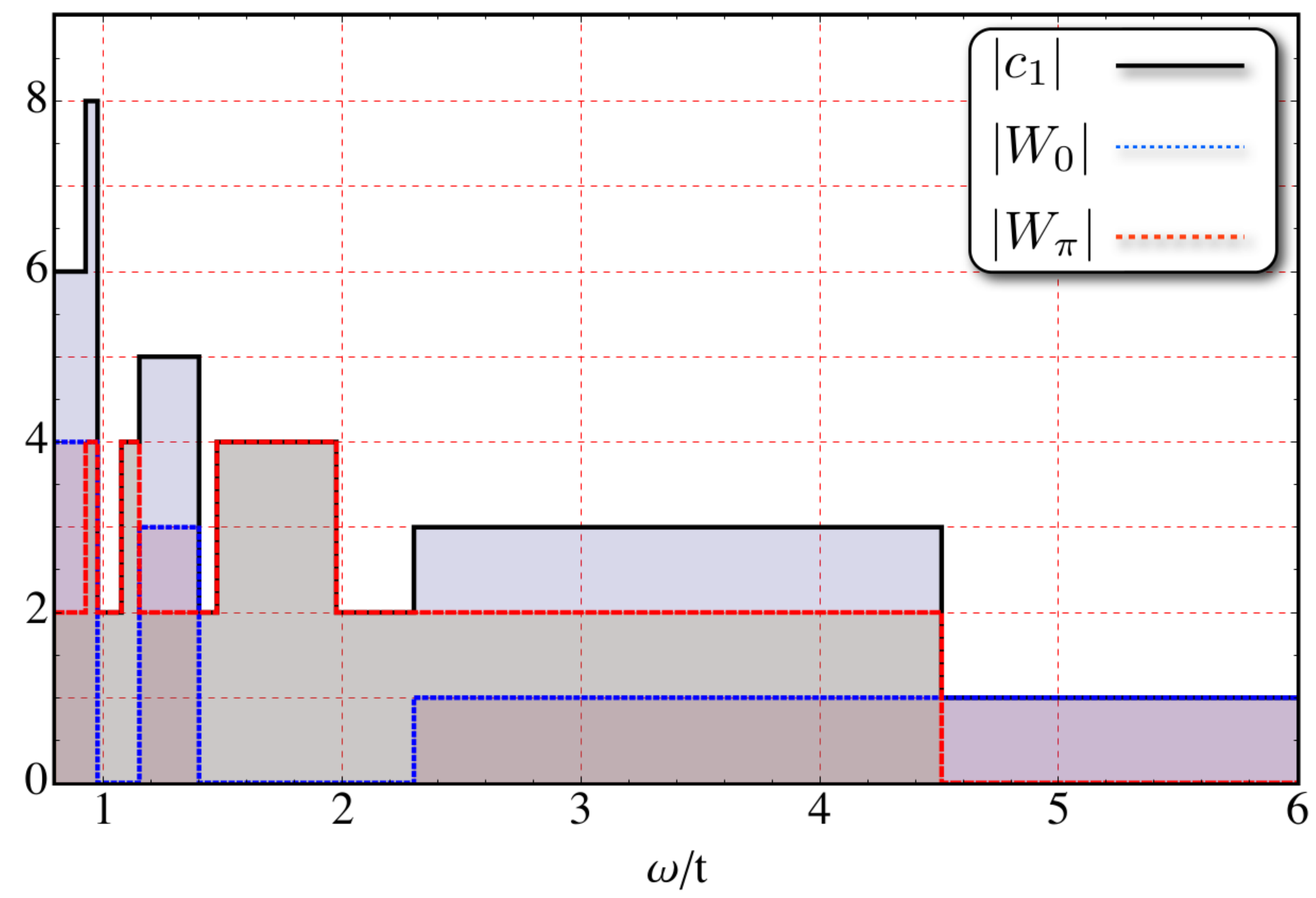} \caption{\label{fig:Plateaus}Chern number and number of chiral edge states in each gap versus the frequency $\omega$.
A change in frequency creates a plateau structure in which each step
is related to a closure of one of the gaps $\Delta_{0,\pi}$.}
\end{figure}
The plateau structure of the Chern number gives information about the number of chiral edge states, but
is not sufficient to fully predict the chirality at each gap.
For that purpose, and for the present two-band model, one needs to know either ($c_1$,$W$) where $W=W_0+W_\pi$, or directly ($W _{0}$,$W_\pi$), as represented in Fig.\ref{fig:Plateaus}.


\textbf{\textsl{Possible experimental realizations:}} 
Different realizations of time periodically driven honeycomb lattices can exhibit Chern number plateaus, as well as anti-chiral edge states.
Shaken optical lattices \cite{HAuke} and photonic crystals with extended helical wave guides in the third spatial dimension (that simulates a circular in plane vector potential \cite{rechtsman13}) constitute concrete physical systems with a high level of control, in which the different phases discussed here can be achieved.
Micro-wave honeycomb crystals have also been demonstrated to be correctly described by a tight-binding model \cite{bellec}.
The recent realization of a Floquet micro-wave crystal exhibiting side-bands features \cite{gehler}, provides an additional promising direction for
the achievement of the different regimes we describe here (the hopping parameter being of the order of a few MHz \cite{bellec} whereas 
the driving frequency of the cavity can vary up to the GHz \cite{gehler}).
In graphene, the possibility to induce phases with higher Chern numbers would correspond to electromagnetic frequencies $\omega/2\pi$ above the THz, from the mid infrared to visible light.
The quasi-energy gaps $\Delta_{0,\pi}$ and the topological edge states predicted in the present work could be probed by time-and-angle-resolved photoemission spectroscopy \cite{Sabota,Hajlaoui,Wang},
as recently demonstrated on the surface of a 3D topological insulator\cite{Gedik}. Also, a direct connection between the edge states in the gaps $\Delta_{0,\pi}$ 
and the quantization of the differential conductance in a two-terminal setup
has been found at high frequency in graphene \cite{FertigPAper} as well as for arbitrary frequencies in driven superconductors exhibiting Floquet Majorana particles \cite{Transport-Signatures,Karch}. 
It is therefore natural to expect a similar signature of the phases we find through the number of chiral and anti-chiral edge states in a DC transport measurement,
but a rigorous treatment of this question is needed and is an important issue for future investigations.


\textbf{\textsl{Summary:}} 
We have proposed a general mechanism to obtain Chern  phases with arbitrary values of the Chern number by periodically driving 2D Dirac semi-metals.
We have performed a direct calculation of the Chern number for arbitrary frequencies. 
Its behavior, as a function of the frequency, yields a plateau structure for the QAHE.
We have also found that the Floquet Chern phases can exhibit anti-chiral edge modes in either of the two relevant gaps.
Our model reveals as well the role of the merging and the creation of Dirac points to generate such modes.
Finally we have suggested possible realizations and observations of these phases in various physical systems.


We acknowledge A. G. Grushin for useful discussions. P. D. was supported
by the European Marie Curie ITN NanoCTM. \'{A}. G\'{o}mez-Le\'{o}n acknowledges
JAE program. \'{A}. G. L. and G. P. acknowledge MAT 2011-24331 
grant 234970 (EU) for financial support.

\begin{widetext}

\title{Supplementary information}
\maketitle

\appendix

\section{Sambe space representation of the Floquet operator}

For the calculation of the Floquet operator we first consider the tight-binding Hamiltonian of the
undriven honeycomb lattice in reciprocal space, within the nearest neighbor
approximation, and for the basis $\left(u_{A,\mathbf{k}},u_{B,\mathbf{k}}\right)^{T}$:
\begin{equation}
H\left(\mathbf{k}\right)=t\left(\begin{array}{cc}
0 & \sum_{j=1}^{3}e^{i\mathbf{k}\cdot\mathbf{a}_{j}}\\
\sum_{j=1}^{3}e^{-i\mathbf{k}\cdot\mathbf{a}_{j}} & 0
\end{array}\right),
\end{equation}
where $\mathbf{a}_{1}=a\left(0,0\right)$, $\mathbf{a}_{2}=a\left(\frac{3}{2},\frac{\sqrt{3}}{2}\right)$,
$\mathbf{a}_{3}=a\left(\frac{3}{2},-\frac{\sqrt{3}}{2}\right)$. The
application of the AC field is included by means of the vector potential
$\mathbf{A}\left(\tau\right)$, as a time dependent phase factor attached
to the hoppings $t\rightarrow t_{j}\left(\tau\right)=te^{i\mathbf{A}\left(\tau\right)\cdot\mathbf{d}_{j}}$,
being $\mathbf{d}_{1}=a\left(1,0\right)$, $\mathbf{d}_{2}=\frac{a}{2}\left(-1,\sqrt{3}\right)$,
and $\mathbf{d}_{3}=\frac{a}{2}\left(-1,-\sqrt{3}\right)$. Thus,
the time dependent Hamiltonian for the periodically driven honeycomb lattice is:

\begin{eqnarray}
H\left(\mathbf{k},\tau\right) & = & t\left(\begin{array}{cc}
0 & \rho\left(\mathbf{k},\tau\right)\\
\rho\left(\mathbf{k},\tau\right)^{*} & 0
\end{array}\right),\\
\rho\left(\mathbf{k},\tau\right) & \equiv & \sum_{j=1}^{3}e^{i\left(\mathbf{k}\cdot\mathbf{a}_{j}+\mathbf{A}\left(\tau\right)\cdot\mathbf{d}_{j}\right)}.
\end{eqnarray}
The Floquet operator is given by $\mathcal{H}\left(\mathbf{k},\tau\right)=H\left(\mathbf{k},\tau\right)-i\partial_{\tau}$,
and its representation in Sambe space, leads to the matrix elements
given in Eq.1 in the main text. They are expressed in terms
of the time-independent basis $\left\{ |u_{\alpha,\mathbf{k},p}\rangle\right\} $,
where $p$ corresponds to the $p$-th coefficient of the Fourier expansion
of $|u_{\alpha,\mathbf{k}}\left(t\right)\rangle$. Note that the importance
of the Sambe space representation is that the time average included
in the composed scalar product $\langle\langle\ldots\rangle\rangle=\frac{1}{T}\int_{0}^{T}\langle\ldots\rangle dt$
allows the use of the time independent basis $\left\{ |u_{\alpha,\mathbf{k},p}\rangle\right\} $
by increasing the matrix dimension. The explicit derivation of the Floquet operator in Sambe space can be seen in Ref.\cite{OURPAPER}.

\section{Derivation of the effective Hamiltonian $H_{\text{eff}}$}

For the derivation of the effective Hamiltonian considered in Eq.5 in the main text, 
we need to assume a description of the long time dynamics ($\tau\gg1/\omega$).
Let us consider a stroboscopic evolution operator for the time periodic
Hamiltonian over a period $T$, and an approximate evolution operator
described by a time independent effective Hamiltonian $H_{\text{eff}}$:
\[
U\left(T\right)=\mathcal{T}e^{-i\int_{0}^{T}H\left(\tau\right)d\tau}\simeq e^{-iH_{\text{eff}}T}.
\]
Then, we can consider a Fourier expansion of the Hamiltonian due to
its time periodicity: 
\begin{equation}
H\left(\tau\right)=\sum_{n=-\infty}^{\infty}H_{n}e^{in\omega\tau}\simeq H_{0}+H_{1}e^{i\omega\tau}+H_{-1}e^{-i\omega\tau},\label{eq:FirstOrderHam1}
\end{equation}
where we have considered just the first harmonic contribution. Next,
expanding the exponential in Taylor series we obtain: 
\[
e^{-i\int_{0}^{T}H\left(\tau\right)d\tau}\simeq1-i\int_{0}^{T}H\left(\tau\right)d\tau+\frac{\left(-i\right)^{2}}{2}\int_{0}^{T}H\left(\tau_{1}\right)d\tau_{1}\int_{0}^{T}H\left(\tau_{2}\right)d\tau_{2}+\ldots
\]
In terms of the previous expansion, the stroboscopic evolution
operator is given by: 
\begin{eqnarray*}
U\left(T\right) & \simeq & \mathcal{T}\left\{ 1-i\int_{0}^{T}H\left(\tau\right)d\tau+\frac{\left(-i\right)^{2}}{2}\int_{0}^{T}H\left(\tau_{1}\right)d\tau_{1}\int_{0}^{T}H\left(\tau_{2}\right)d\tau_{2}\right\} \\
 & = & 1-i\int_{0}^{T}H\left(\tau\right)d\tau-\frac{1}{2}\left[\int_{0}^{T}d\tau_{1}\int_{0}^{\tau_{1}}d\tau_{2}H\left(\tau_{1}\right)H\left(\tau_{2}\right)+\int_{0}^{T}d\tau_{2}\int_{0}^{\tau_{2}}d\tau_{1}H\left(\tau_{2}\right)H\left(\tau_{1}\right)\right],
\end{eqnarray*}
where in the last line we have applied the time ordering operator
$\mathcal{T}$. Using Eq.~\ref{eq:FirstOrderHam1} we can finally perform
the integrals above, and obtain the effective evolution operator:
\begin{eqnarray*}
U\left(T\right) & \simeq & 1-iH_{0}T-\frac{T}{\omega}\left\{ \pi H_{0}^{2}-i\left(\left[H_{0},H_{-1}\right]-\left[H_{0},H_{1}\right]+\left[H_{-1},H_{1}\right]\right)\right\} \\
 & \simeq & 1-iH_{\text{eff}}T-\frac{1}{2}H_{\text{eff}}^{2}T^{2}+\ldots
\end{eqnarray*}
By direct comparison the effective Hamiltonian reads: 
\begin{equation}
H_{\text{eff}}=H_{0}-\frac{1}{\omega}\left(\left[H_{0},H_{-1}\right]-\left[H_{0},H_{1}\right]+\left[H_{-1},H_{1}\right]\right),\label{eq:Effective}
\end{equation}
where we have neglected the term $H_{0}^{2}$ because it belongs to
the second order term of the Taylor expansion. It is worth to mention that we have also derived  Eq.~\ref{eq:Effective}
by using the more rigorous Magnus expansion.
Note that an expansion to next order yields terms which are not proportional to the Pauli matrix $\sigma_z$, and is therefore useless for our purposes. In addition, the gap is developed just when the $\sigma_z$ component of the Hamiltonian breaks TRS. This will happen when $\phi\neq 0,\pi$, since the phases attached to the hoppings in Sambe space (Eq.2) are non zero. This makes the Hamiltonian to become a complex matrix, and removes the accidental degeneracies (Dirac cones) that were present in the absence of complex phases.
It must also be noticed that this result is similar to the one obtained for
example in \cite{DemlerPRBGRaphene}. Nevertheless, as we did not require a
low energy approximation to a single Dirac cone, we are not restricted
to small values of the field amplitude. This feature allows us to fully address
the phase diagram and the dependence of the Chern number on the external
parameters.

\section{Edge states}

Here we describe the properties of the different edge states appearing
in our system, depending on the regime under consideration. In the high
frequency regime, the Floquet side-bands are far separated to each other, then the
gap $\Delta_{\pi}$ is always trivial and does not develop topologically
protected edge states (in analogy with what is usually referred as the atomic limit).
Thus, all topological properties can be studied within a two-band
approximation and the zero energy modes behave equivalently to the
ones of static systems. This is plotted in Fig.\ref{fig:Edge-states-HF}
for zigzag (a) and armchair (b) ribbons, including a color code for the time average density distribution at the edges: $\Delta\bar{\rho}=\bar{\rho}(\mathbf{r}_0)-\bar{\rho}(\mathbf{r}_N)$.

\begin{figure}[H]
\begin{centering}
\includegraphics[scale=0.45]{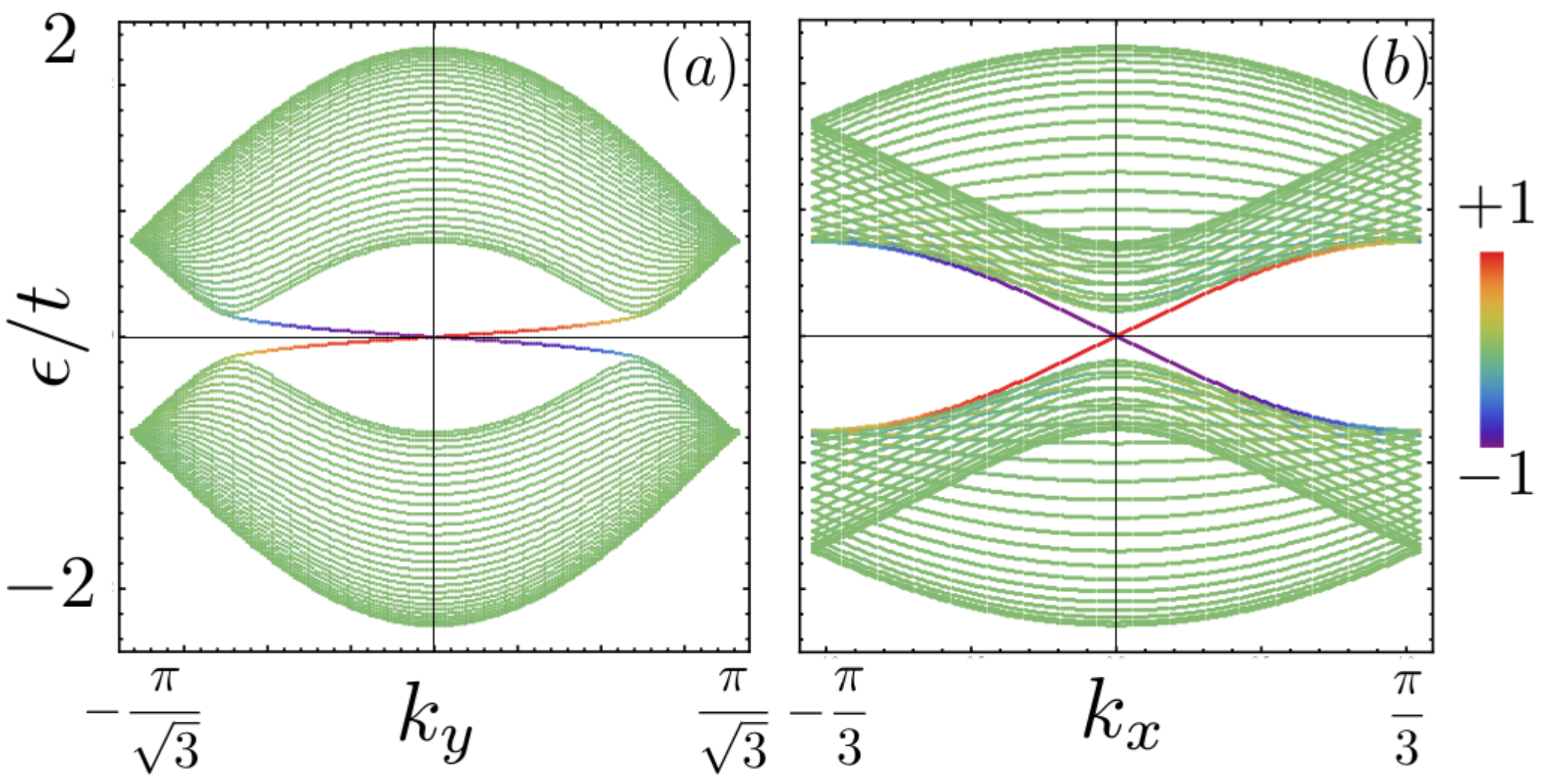} 
\includegraphics[scale=0.45]{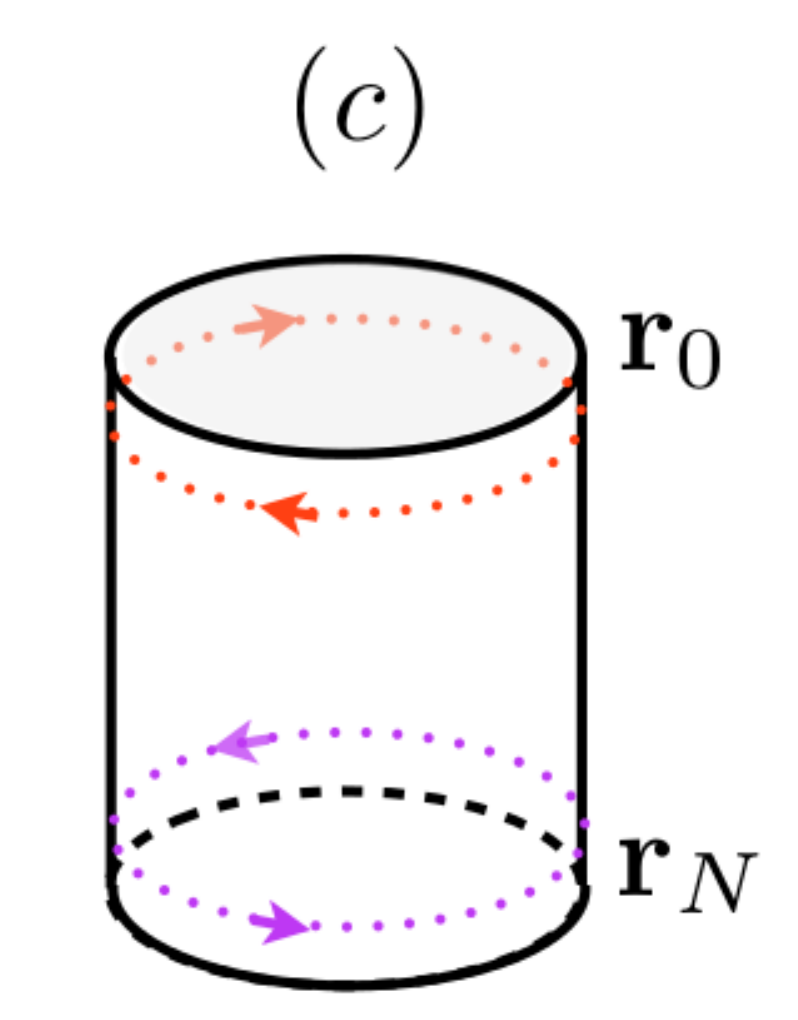}
\par\end{centering}

\caption{\label{fig:Edge-states-HF}Quasi-energies and edge states for
the high frequency regime ($\omega=10t$). (a) shows the zigzag ribbon
spectrum with two chiral edge states propagating along the top/bottom
(red/blue) boundary while (b) shows the armchair spectrum. In this
regime $\Delta_{\pi}$ corresponds to a trivial gap, and the edge states
 can only cross $\Delta_{0}$. (c) shows an schematic figure of the edge states propagation. Parameters considered: 30 unit cells
of length for the ribbons, $t=1$, $\phi=\pi/2$ and $A_{x}a=A_{y}a=1$.}
\end{figure}

In addition, in Fig.\ref{fig:chirality} we plot the edge states chirality at each boundary for different field amplitudes and fixed phase difference $\phi=\pi/4$. 
It shows how it is possible to reverse the chirality at each boundary by just tuning the amplitude of the field, annihilating and creating a new pair of Dirac points with opposite topological charge.

\begin{figure}[H]
\begin{centering}
\includegraphics[scale=0.4]{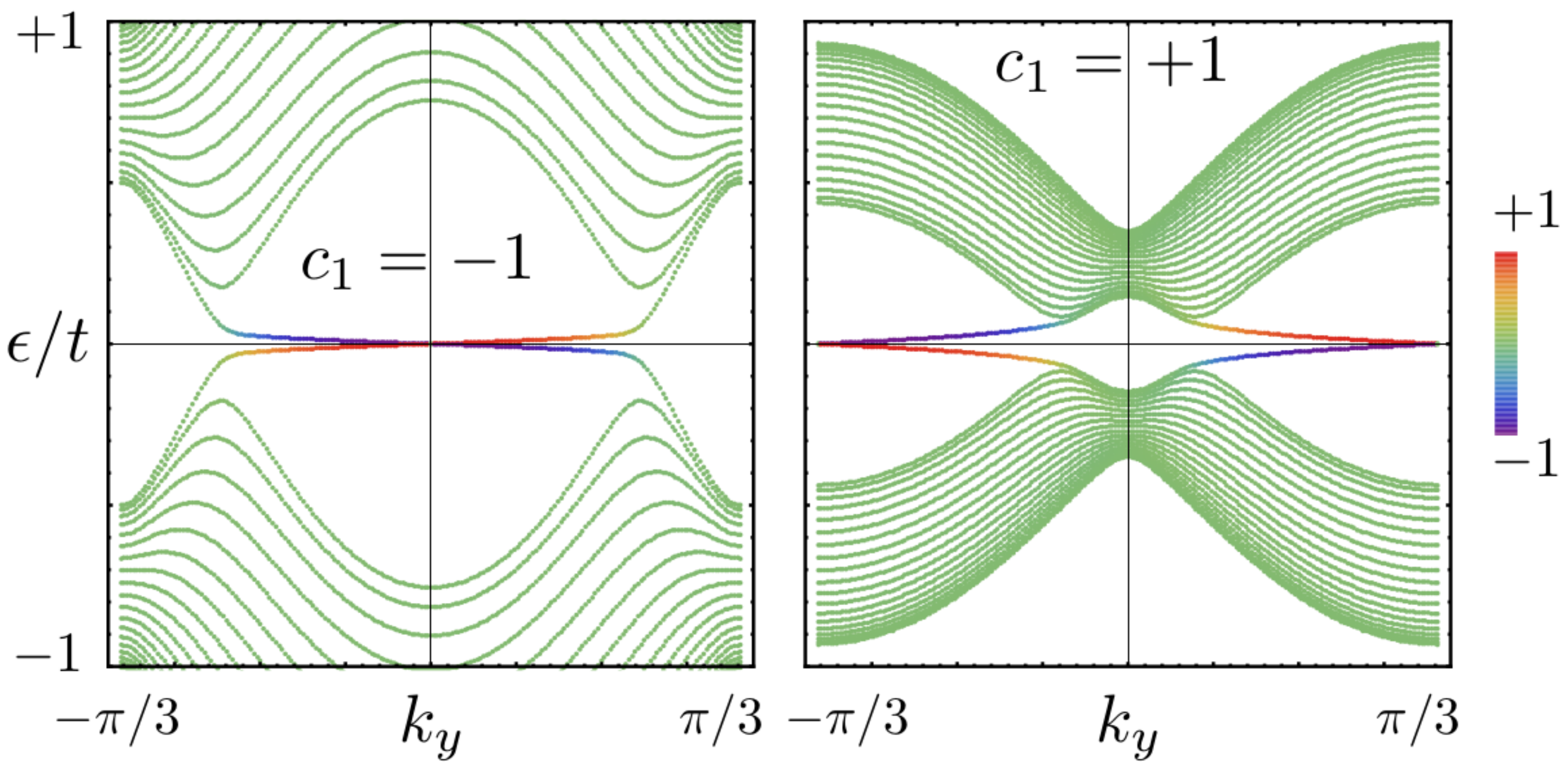} 
\par\end{centering}

\caption{\label{fig:Edge-states-chirality}Edge states for a zigzag ribbon at 
different field amplitudes $A_xa=A_ya=1,3$ (left/right repectively) in
the high frequency regime ($\omega=10t$). The left plot shows 
two chiral edge states propagating along the top/bottom
(red/blue) boundary, while the right plot shows the same edge states, 
moving in the opposite direction. Parameters considered: 20 unit cells
of length for the ribbon and $\phi=\pi/4$.}
\label{fig:chirality}
\end{figure}

On the other hand, when the frequency is decreased up to a certain $\omega$
value, the gap between different side-bands $\Delta_{\pi}$ collapses. This exactly 
happens when the frequency reaches the band width value for some fixed $A_{x,y}$.
 When the gap $\Delta_{\pi}$ is reopened the apparition of new
edge states is possible, as we show in Fig.3 in the main text.
For the case of the honeycomb lattice, the gap $\Delta_{\pi}$ develops
two pairs of chiral edge states connecting different side-bands, and
in consequence the Chern number changes by two units.

To conclude, when the system has a Chern number
$c_{1}=0$, chiral edge states do not appear (see Fig.\ref{fig:Edgestates2-1}).
Instead, the existence of localized end states is restricted by whether
the system is in a Zak insulating phase or not. Note that
these states are disconnected from the bulk and
they are not dispersive. The edge states arising from a non zero Chern
number phase cross the gap, and represent
a condensed matter realization of the Haldane model, where time reversal
is broken for zero magnetic field.

\begin{figure}[H]
\begin{centering}
\includegraphics[scale=0.4]{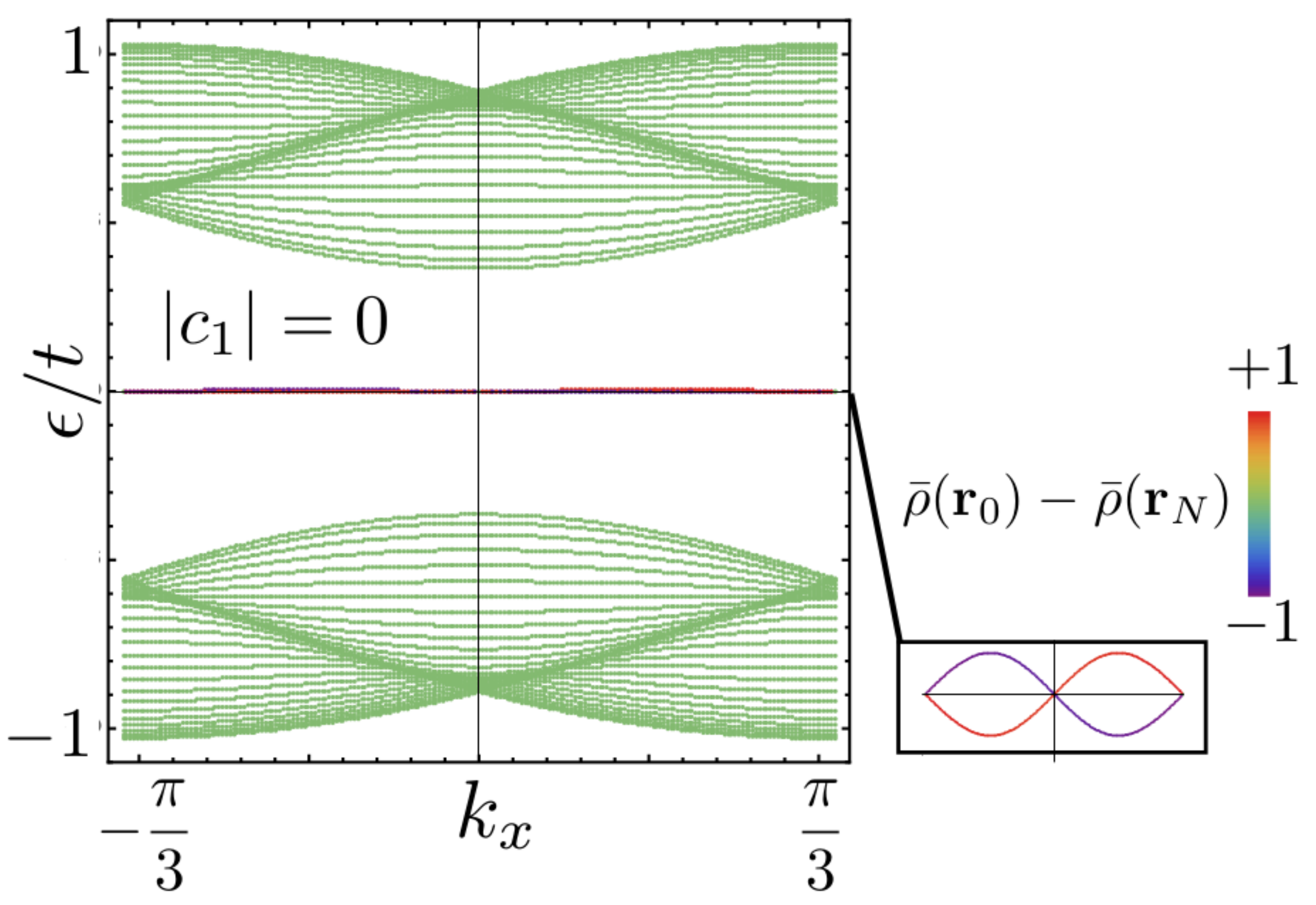} 
\par\end{centering}

\caption{\label{fig:Edgestates2-1}Quasi-energy spectrum for graphene armchair
ribbon in a topologically trivial region. The parameters are: $A_{x}a=A_{y}a=2$,
$\phi=0.6$, $\omega=10t$, and 30 unit cells for the ribbon. For
a zigzag ribbon non topologically protected edge states are also
found (not shown).}
\end{figure}

Finally in Fig.\ref{fig:Full-numerical-calculation} we plot the full
plateau structure obtained numerically in terms of the Chern number. Note that the inequivalent bands within a single Floquet side-band in a periodically
driven two-band system necessarily have opposite Chern numbers\cite{RUdnerPRX},
so that we only need to compute the Chern number of one of them. In addition, note that the increasing number of edge states is also coincident with the one obtained in other periodically driven systems, such as in Ref.\cite{MajoranaSteps}.

\begin{figure}
\begin{centering}
\includegraphics[scale=0.35]{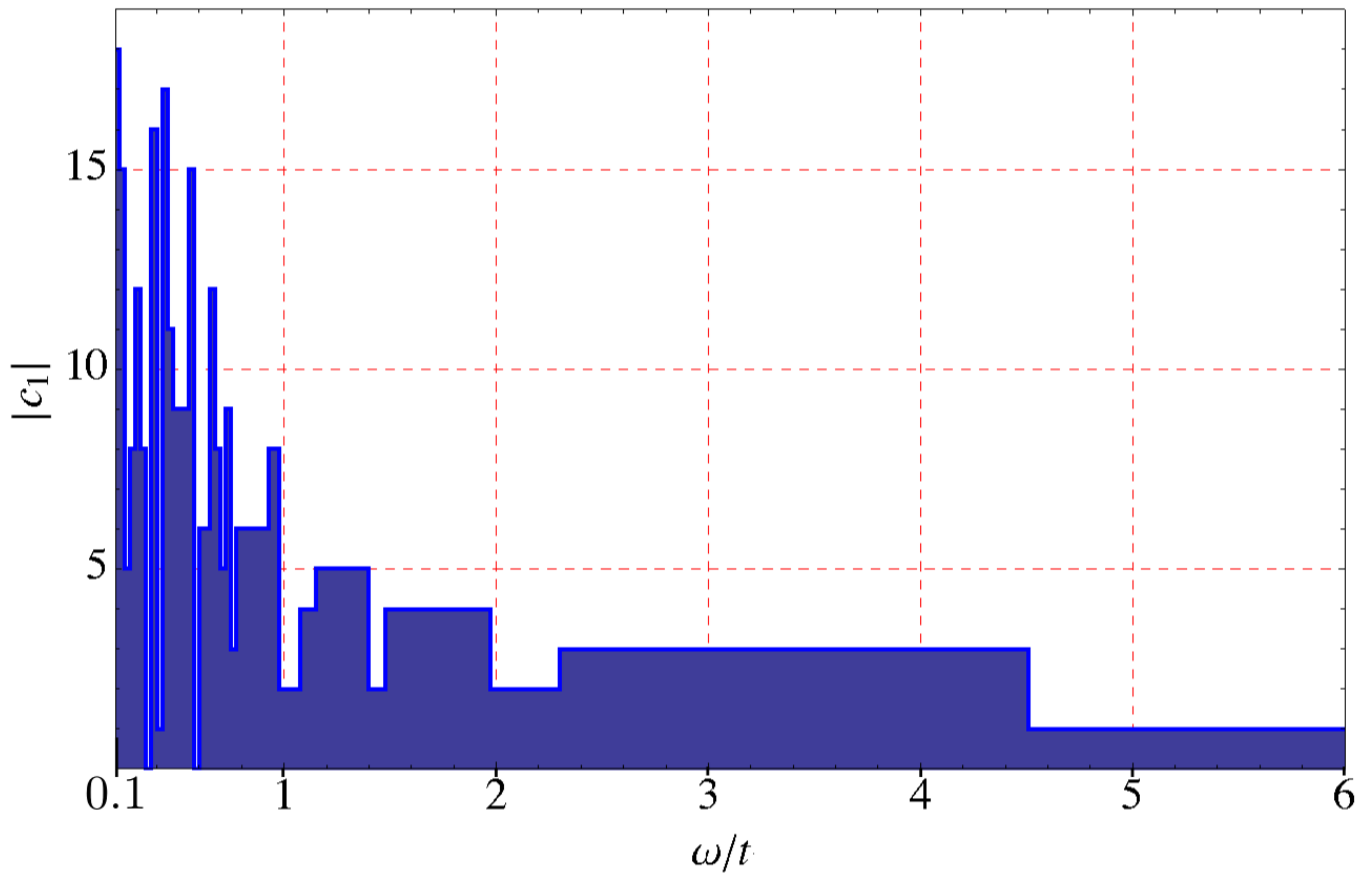} 
\par\end{centering}

\caption{\label{fig:Full-numerical-calculation}Full $\omega$ range of the numerical calculation
for the Chern number (absolute value) versus the frequency in units of the hopping energy $\omega/t$. Note that the Chern
number increases with the successive gap closures. The parameters considered are: $A_x a=A_y a=1$, and $\phi=\pi/2$.}
\end{figure}

\end{widetext}

\end{document}